\documentclass[prl,aps,showpacs,groupedaddress,superscriptaddress,twocolumn]{revtex4-1}
\usepackage{amsmath,amssymb}
\usepackage[utf8x]{inputenc}
\usepackage[english]{babel}
\usepackage{graphicx,epsfig}
\usepackage{todonotes}
\usepackage{hyperref}

\usepackage{braket}
\usepackage{soul}
\usepackage{color}
\makeatother
\usepackage{placeins}

\usepackage{tikz}
\usetikzlibrary{positioning,arrows}
\usetikzlibrary{decorations.pathmorphing}
\usetikzlibrary{decorations.markings}
\usetikzlibrary{calc}
\usetikzlibrary{patterns}
\usetikzlibrary{shapes.misc}
\tikzset{
phys/.style={thick, postaction={decorate}, decoration={markings, mark=at position .7 with {\arrow[]{triangle 45}}}},
res/.style={thick, decorate, decoration={snake,segment length=7, amplitude=1.5}},
arrow/.style={thick, draw=white, postaction={decorate}, decoration={markings, mark=at position .6 with {\arrow[black]{triangle 45}}}}
 }

\begin{document}

\title{Universality class of Ising critical  states with long-range losses}
 
\author{Jamir Marino}
\affiliation{Institut f\"ur Physik, Johannes Gutenberg Universit\"at Mainz, D-55099 Mainz, Germany}
\affiliation{Kavli Institute for Theoretical Physics, University of California, Santa Barbara, CA 93106-4030, USA}

\date{\today}

\begin{abstract}
 
{We show that spatial resolved dissipation can act on $d$-dimensional spin systems in the Ising universality class  by qualitatively modifying   the nature of  their critical points.
 We consider  power-law decaying spin losses with a   Lindbladian spectrum  closing at small momenta   as $\propto q^\alpha$, with $\alpha$ a positive  tunable exponent directly related to the power-law decay of the spatial profile of   losses at long distances, $1/r^{(\alpha+d)}$.
 This  yields a   class of   soft modes asymptotically decoupled from dissipation at small momenta, which are responsible} for the emergence of   a critical scaling regime ascribable to the non-unitary counterpart of   the   universality class of  long-range interacting  Ising models. 
For $\alpha<1$  we find a non-equilibrium critical point ruled by a dynamical field theory described by a Langevin model with coexisting inertial ($\sim {\partial^2_t}$) and frictional ($\sim {\partial_t}$) kinetic coefficients, and driven by a gapless Markovian noise with variance $\propto q^\alpha$ at   small momenta. 
This  effective field theory is beyond the Halperin-Hohenberg description of dynamical criticality, and   its critical exponents differ from their  unitary long-range counterparts. 
 %
 Our work lays out   perspectives for a revision of    universality   in driven-open systems by employing dark states taylored by   programmable dissipation.

\end{abstract}
\maketitle

%

\paragraph{Introduction --} 
The search for non-equilibrium criticality in driven open quantum systems has  become an exciting research frontier, both for its fundamental relevance in statistical mechanics, and for the variety of AMO platforms   where it can be concretely explored~\cite{hartmann08,Syassen2008,Yamamoto10,baumann2010dicke,Weitz10,blatt12,britton12,houck12,Carusotto13,Carr13,Zhu14,klinder2015dynamical,fitzpatrick2017observation,dogra2019dissipation,kessler2020observation,ferri2021emerging,nagy2011critical,nagy2015nonequilibrium,lundgren2020nature}.
The aim is the  discovery of universality classes which cannot be encompassed by established  classifications of dynamical criticality~\cite{Hohenberg1977, Sondhi1997, vojta2003quantum, sachdev2007quantum} nor  can be related to out-of-equilibrium     scaling in isolated systems~\cite{orioli2015universal,berges2015universality, prufer2018observation,erne2018universal,eigen2018universal,schmied2020non, glidden2021bidirectional,chiocchetta2017dynamical,sciolla2010quantum,schiro2010time,eckstein2009thermalization,vzunkovivc2018dynamical,halimeh2017prethermalization}, where, in sharp contrast, both energy and total number of particles are conserved. 
A common obstruction against the realization of this program is the occurrence of an effective thermal behaviour for the   soft modes relevant at the critical points of driven-dissipative systems~\cite{mitra06,PhysRevB.85.184302, DallaTorreDiehl2013, maghrebi2016nonequilibrium, foss2017emergent, kirton2019introduction}: 
although  the full momentum distribution of the non-equilibrium steady state manifestly breaks detailed balance, low momenta can thermalize at an effective temperature set by the interplay of drive, noise and losses. This forces several instances of driven-open criticality  to fall into known equilibrium universality classes~\cite{Sieberer2016review,Hohenberg1977}, with few exceptions represented by the appearance of novel    independent anomalous exponents associated to decoherence~\cite{Sieberer2013, Sieberer2014,Tauber2014}, or by exotic features as non-equilibrium multi-critical points~\cite{young2020nonequilibrium}.

The culprit for   effective thermalization   is a  noise variance (dictated by dissipation)  with a non-vanishing gap at small momenta and/or frequencies, which sets    the   temperature of infrared modes in several circumstances of interest~\cite{mitra06,PhysRevB.85.184302, DallaTorreDiehl2013, maghrebi2016nonequilibrium, foss2017emergent, Sieberer2013, Sieberer2016review,wald2018lindblad}. Softening such gap and allowing the noise   to scale down to zero at small momenta,   is the route for   instances of driven-open criticality without thermal counterpart. In quadratic fermionic models~\cite{proseneisert,Moos} or interacting quantum wires~\cite{dalla2010quantum, Marino2016,baldwin2021singularities},   dissipation with non-local support in real space   acting on neighboring sites in a correlated fashion~\cite{diehl2008quantum}, has been employed to achieve non-equilibrium quantum criticality.  In these cases the     noise  variance
vanishes at  infrared momenta, and it   exposes a set of   modes   asymptotically  decoupled for $q\to0$ from the decohering and thermalizing effect of the environment. These forms of non-local dissipation can steer a  system into a many-particle dark state with non-trivial    quantum correlations --  a state preparation protocol with  interesting perspectives for applications in quantum information and technology~\cite{eisert2010noise,verstraete2009quantum,marcos2012photon,diehl2011topology,  PhysRevA.86.013606,buvca2019non,barreiro2011open,lin2013dissipative,ma2019dissipatively,PhysRevA.89.023616,reiter2016scalable,parmee2018phases}.

 In this work, we consider spin losses with a  controllable spatial profile decaying algebraically at long distances~\cite{gonzalez2015subwavelength,hung2016quantum,daley,kushal}.  
Their Lindbladian spectrum scales with momentum softly as $\propto q^\alpha$ in the infrared; the tunability of $\alpha$ allows us to explore a dissipative analogue of the universality class of  long-range interacting quantum magnets. Our results are based on  renormalization group (RG)   and therefore pertinent to a whole family of spin models  distinct by {RG} irrelevant perturbations at the Ising critical point. {Modern cavity QED quantum simulators~\cite{hung2016quantum,PhysRevLett.123.130601,MonikaPRL2019,periwal2021programmable} in the regime of strong cavity loss, have the potential to expose uncoventional forms of dynamical criticality, since they can imprint on atomic ensemble decay channels with   tunable spatial profiles~\cite{kushal}. This is in sharp contrast with  previous contributions on driven open  criticality where the structure of  dissipation supporting dark modes is not flexible and given by the specific implementation at hand~\cite{eisert2010noise,marcos2012photon,  PhysRevA.86.013606, Marino2016}. In particular, we  discuss here the instance of critical spin ensembles  subject to  long-range spatial emission, whose universal properties are} ruled by a Langevin theory~\cite{Hohenberg1977,Tauber2014} where inertial ($\propto {\partial^2_t}$) and frictional ($\propto {\partial_t}$) kinetic coefficients  coexist and with a gapless driving noise    scaling proportionally to $q^\alpha$ in the infrared. Upon tuning $\alpha$, one can control  the degree of RG relevance of the operators necessary for a consistent description of these novel critical states, and interpolate among different universality classes.\\
 
\paragraph{Ising criticality with non-local losses -- } 
We consider a quantum Ising chain in $d$ dimensions 
\begin{equation}\label{Ising}
H=-\sum_{\langle i,j\rangle} \sigma^x_i \sigma^x_j +h\sum_i\sigma^z_i,
\end{equation}
subject to a spin loss Lindblad channel which is non-local  and   shaped by a spatial structure factor $\gamma_{i,j}$:
\begin{equation}\label{eq:model}
	\resizebox{.896\hsize}{!}{	$\dot{\rho}=i[\rho,{H}]
	+\sum_{i,j}\gamma_{i,j}\left(\sigma^-_{i}\rho\sigma_{j}^{+}-\frac{1}{2}\left\{ \sigma_{j}^{+}\sigma^-_{i},\rho\right\} \right)$}.
\end{equation}
{The open quantum system in~\eqref{Ising} and~\eqref{eq:model} is not exactly solvable and, even with state-of-art numerics, dynamics could be extracted only for small system sizes and intermediate times. Instead, here we rely on   non-equilibrium RG  to inspect the long-distance/long-time scaling  properties of the system at criticality.} In this regard, any {RG} irrelevant perturbation at the Ising critical point in~\eqref{Ising} (e.g. short-range spin-spin interactions along the $\hat{z}$ direction) will not affect our results, which are therefore pertinent to the whole set of spin models belonging to the Ising universality class. Correlated spin losses as in~Eq.\eqref{eq:model} can be realized in cavity QED~\cite{kushal} or photonic crystal waveguides~\cite{gonzalez2015subwavelength,hung2016quantum}, where tunable interactions  and    losses between pairs of spins at arbitrary distances can be controlled through a combination of spatial-dependent   energy level shifts and external pump fields~\cite{kushal}. The former typically result from a background magnetic field gradient coupling to  levels in hyperfine manifold, and they gives spatial resolution to dissipation, while a Raman drive with several sidebands enables control on the functional form of $\gamma_{i,j}$, which can be   made translational invariant, $\gamma(r)$ with $r=|i-j|$. 
\begin{figure}[b!]
	\includegraphics[width=7cm]{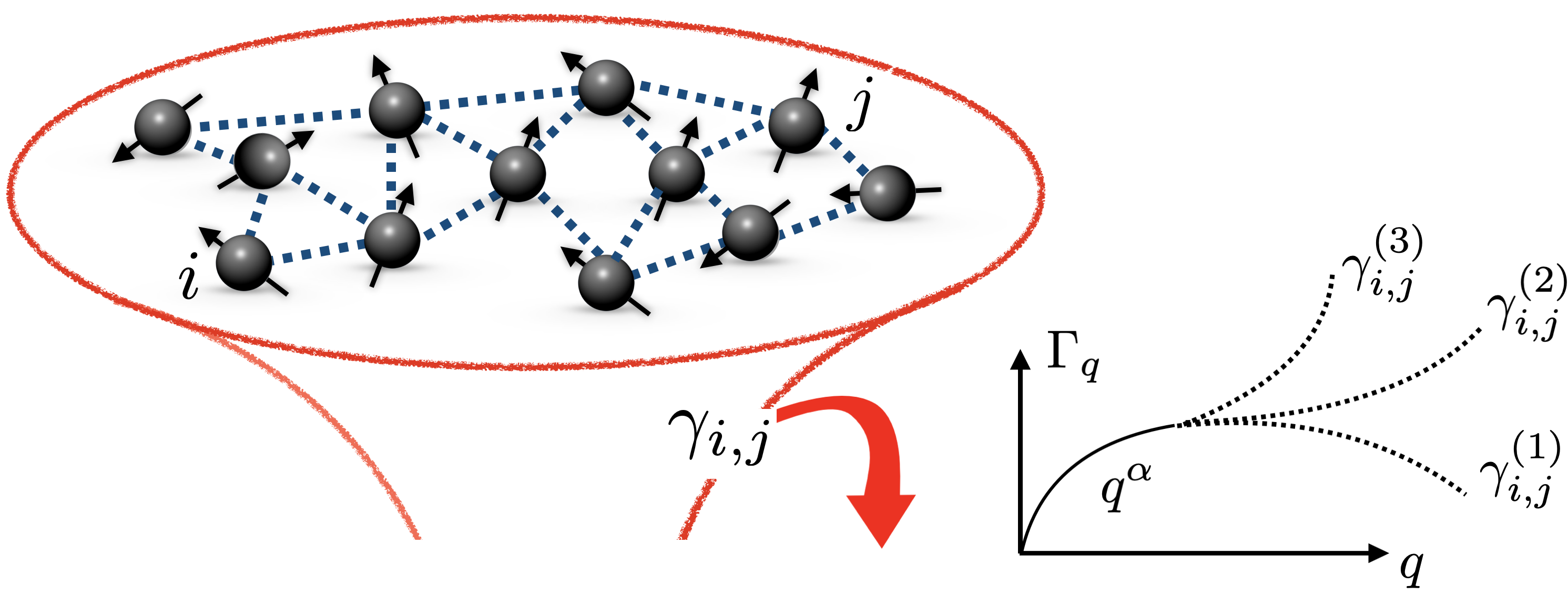}
	\caption{Schematic portrait of a spin lattice   subject to non-local losses, $\gamma_{i,j}$, acting on pairs of spins at positions $i$ and $j$.
	%
	%
	%
	}\label{fig1} 
\end{figure}

In the following we   analyze the impact of a   class   of $\gamma(r)$,  {which scales asymptotically as an inverse power law of $r$,  on Ising critical systems (i.e. $\gamma(r)\propto 1/r^{\alpha+d}$ for large $r$, cf. with Fig.~\ref{fig1}).  The details of the implementation of $\gamma(r)$ are contained in Refs.~\cite{hung2016quantum,kushal} } (see also    note~\footnote{ The relation between $\Gamma_q$ and the amplitudes, $\Omega_b$, of the Raman sidebands is $\Omega_b\sim\sum_q e^{iqb}\sqrt{\Gamma_q}$ (see Ref.~\cite{kushal}). In order to have $\Gamma_q\sim q^\alpha$ for $q\to0$, the amplitudes have to scale as $\Omega_b\sim b^{-(d+\alpha/2)}$ for $b\gg1$.}). { Diagonalizing Eq.~\eqref{eq:model}  in Fourier space with this shape   of $\gamma(r)$, we obtain for the  decay rate of the infrared modes the expression $\Gamma_q\sim \Gamma_1 q^\alpha$. These infrared modes are dark},  i.e. asymptotically decoupled from dissipative effects ($\Gamma_q$ vanishes as $q\to0$).  
Following the analogy with criticality in long-range interacting spin systems where instead  the Hamiltonian spectral gap   scales as $m\sim q^\alpha$~\cite{fisher1972critical,dutta2001phase,maghrebi2016causality}, we can interpret the instance of criticality inspected in this work as a non-unitary counterpart of critical long-range interacting Ising systems. This analogy is further supported by the fact that the anti-commutator in the Lindblad equation~\eqref{eq:model} can be   regarded  as a non-hermitian long-range interaction term~\cite{kushal}. 
We notice that the shape of $\gamma(r)$ at short or intermediate distances is   inconsequential for our results which are   relevant  for  soft, infrared modes  at criticality. Therefore, our analysis     applies to a broad class of  spatial profiles $\gamma(r)$, provided they entail the infrared scaling of the Lindbladian spectrum  mentioned above, $\Gamma_q\sim\Gamma_1q^\alpha$ (cf. with the inset in Fig.~\ref{fig1}). 
In this respect, our setup also extends previous instances on   preparation and criticality of open quantum systems with dark states in cold atoms or quantum optics platforms, where, in contrast, the shape of non-local losses is not tunable~\cite{proseneisert,Moos,marcos2012photon,dalla2010quantum, Marino2016}. For the case discussed in this work, the exponent $\alpha$   can be flexibly varied by a proper choice of the amplitudes of the Raman sidebands (see note [73], or for more details Ref.~\cite{kushal}).
 {Notice that the $\alpha=0$ case will not   display any interesting instance of dissipative criticality since it does not support dark states ($\Gamma_q$   constant for $q\to0$)}.\\ 

\paragraph{Canonical scaling with long-range losses -- } 
We now map the lattice model in Eq.~\eqref{eq:model} into a long wavelength non-equilibrium field theory~\cite{Kamenevbook2011,Sieberer2016review}. 
In particular, we will discuss how the effective field theory governing non-equilibrium critical behaviour for $\alpha<1$ is ascribable to a Langevin model~\cite{Hohenberg1977,Tauber2014} with coexisting inertial and frictional terms,   driven by a gapless noise $\propto q^\alpha$ at small momenta.
 Following the usual prescription~\cite{Cardy:1996xt,amit2005field,kardar2007statistical} we map the spin operators in terms of bosons $\sigma^-_i\to a_i$ and $\sigma^z_i\to2a^\dagger_ia_i$, and we implement the hard-core constraint with a large on site non-linearity $a^\dagger_ia^\dagger_ia_ia_i$. By taking the continuum limit and coarse graining over short wavelengths~\cite{Cardy:1996xt,amit2005field,kardar2007statistical,maghrebi2016nonequilibrium}, the Ising interaction in Eq.~\eqref{Ising}   yields a  second derivative in space within a leading order derivative expansion ($Kq^2$ in momentum space), while the non-linearity yields the usual $\varphi^4$  potential. As detailed in Refs.~\cite{lee2013unconventional,maghrebi2016nonequilibrium,Sieberer2016review}, the quantum master equation for an Ising model with losses~\eqref{eq:model} can be mapped into a Keldysh path integral in terms of the classical and quantum components of the real field, $\varphi_{c/q}(Q)$, which in Fourier space, $Q=(\mathbf{q},\omega)$, reads
 \begin{equation}\label{gauss}
 \mathcal{S}_G=\int_{Q} \left( \varphi_c (-Q),\varphi_q(-Q)\right) \begin{pmatrix}
    0 & P^A \\
   P^R & P^K
  \end{pmatrix}
  \begin{pmatrix}
    \varphi_c (Q) \\ \varphi_q(Q)
  \end{pmatrix}
 \end{equation}
 with 
  
  \begin{equation}\label{eqP}\begin{split}
   P^{R/A}&=  -\omega^2 -2Kq^2\mp2i \Gamma_q\omega+m-\Gamma^2_q/2,\\
   P^K&= i\Gamma_q\equiv i(\Gamma_0+\Gamma_1 q^\alpha),
  \end{split}
 \end{equation}
    the retarded/advanced and Keldysh inverse Green's functions~\cite{Kamenevbook2011}. 
  {The former contain  spectral information on the excitations: in our case,   the distance from the critical point, $m$, and the decay rate of the infrared modes, $\Gamma_q$. The Keldysh component of the quadratic action, $P^K$, is instead directly related to the momentum Fourier transform  of the noise variance~\cite{Sieberer2016review}. }
    
    In   $\Gamma_q$ 
 we have included a constant term $\Gamma_0$ which takes into account the spontaneous emission of the spins into free space and which is often the main adversary in schemes implementing dissipative engineering with dark states~\cite{PhysRevA.66.022314,PhysRevLett.110.120402}.  Its detrimental role is to locally measure the atoms and therefore suppress the entangling effect of non-local dissipation at long times (or small wave-vectors). 
 We will inspect the   critical properties of our model in the regime where $\Gamma_1 q^\alpha\gg\Gamma_0$ and therefore $P^K$  scales effectively as $q^\alpha$, and, at the same time, also the retarded/advanced sectors become gapless, $P^{R/A}\sim q^{2\alpha}$, since we tune the spectral mass to zero as well. The former can be implemented in the RG scheme via   the scaling ansatz $\Gamma_0\sim\Gamma_1q^\alpha$ (see also Ref.~\cite{PhysRevB.94.085150}).
As pointed out in~\cite{paz2019driven, paz2021time, paz2021driven}   weak dissipation   can expose novel critical behaviour for a long temporal window before thermalizing effects set it. In our setup, this is mirrored by the fact that for $q\ll (\Gamma_0/\Gamma_1)^{1/\alpha}$ incoherent emission takes over, and the Gaussian action in~\eqref{gauss} reduces to  a Langevin action with $\Gamma_q\simeq\Gamma_0$ and no $\omega^2$ term which notoriously thermalizes~\cite{Hohenberg1977,Tauberbook2014}.    This crossover is analogous to the suppression of equilibrium quantum criticality at distances larger than the de Broglie thermal length~\cite{Sondhi1997, vojta2003quantum, sachdev2007quantum}.

\begin{table*}[t!]
\begin{tabular}{|l|l|l|l|}
\hline
 &  Critical exponent & $d_l<d<d_u$  & Effective field theory  \\ \hline
 LR losses with $\alpha<1$& $\nu=1/(2\alpha)+\epsilon/(12\alpha^2)$  & $2\alpha<d<4\alpha$  & coexisting inertial/frictional derivatives + soft noise  \\ \hline
LR losses with $\alpha>1$ &  $\nu=1/2+\epsilon/12$ & $3-\alpha<d<5-\alpha$ & short-range Ising model + soft noise  \\ \hline
 LR losses and interactions  ($\alpha<2$) & $\nu=1/\alpha+\epsilon/(3\alpha^2)$ &  $\alpha/2<d<3\alpha/2$& long-range Ising model + soft noise  \\ \hline
\end{tabular}
\caption{Non-equilibrium criticality with long-range (LR) losses. The third column displays the lower ($d_l$) and upper ($d_u$) critical dimensions, while the fourth one summarizes the effective dynamical field theory   valid at criticality. {Below the lower critical dimension   separate scaling and RG analyses are required, and an Ising-type field theory description does not apply. The thresholds $d_l$ and $d_u$ implicitly bounds  the   values of $\alpha$ compatible with the universality class discussed in this work. }}\label{tabella}
\end{table*}

 Approaching criticality ($m\to0$) and for $\Gamma_0\to0$, we can adopt the following canonical scaling ansatz~\cite{Tauber2014,ZinnJustinbook} {for the dynamical critical exponent $z$ controlling the relative scaling of frequency and momentum~\cite{Tauberbook2014}}: $\omega\sim q^z$, with $z=\alpha$. This results in the terms $\propto\omega\Gamma_q$ and $\propto\omega^2$ both equally scaling like $\sim q^{2\alpha}$  in the infrared.   %
This is contrast to relaxational Langevin models, where    the inertial term proportional to the second derivative in time ($\sim\omega^2$) is subleading    compared to the frictional first order time derivative  ($\sim\omega$). 
Therefore, we recover a scalar dynamical field theory  with   coexisting inertial and frictional kinetic coefficients,   driven by a gapless Markovian noise, which is a model beyond the  Halperin-Hohenberg classification~\cite{Hohenberg1977}. {An effective field theory resembling some of these features has recently appeared in~\cite{paz2019driven,paz2021driven}. As non-trivial extension here we encompass a family of RG fixed points upon tuning the exponent $\alpha$ of the soft Langevin noise. This results in  corrections not only to dynamical critical exponents as in \cite{paz2019driven,paz2021driven} but also to static ones, as we discuss in the following. }
 
We now focus on kinetic coefficients proportional to spatial derivatives. At the level of canonical power counting, there is  a threshold value    $\alpha<1$ at which the second derivative in space ($\propto Kq^2$ term) is subleading in the infrared compared to the $\Gamma^2_1q^{2\alpha}$ fractional derivative resulting from 'long-range' losses  in the spectral sector ($P^{R/A}$). This should be contrasted with critical long-range interacting Ising models where such threshold is set at    $\alpha=2$~\cite{fisher1972critical,dutta2001phase,vodola2014kitaev, maghrebi2016causality} besides   small corrections resulting from  anomalous dimensions~\cite{sak1973recursion, luijten2002boundary,angelini2014relations, defenu2015fixed,fey2016critical, defenu2017criticality,behan2017long}. These different thresholds occur because hermitian long-range interactions compete with $Kq^2$ through  a $ q^\alpha$ term in the R/A sector (see~\cite{fisher1972critical,dutta2001phase,vodola2014kitaev, maghrebi2016causality}), while non-hermitian ones through  $ q^{2\alpha}$ terms resulting from $\omega\Gamma_q$  and $\Gamma^2_q$   (cf.   with Eq.~\eqref{eqP}). 
We notice that the RG procedure   generates only analytical terms and thus cannot renormalize the  terms scaling with the exponent   $\alpha$ (see also~\cite{fisher1972critical,dutta2001phase, maghrebi2016causality}). The only term which can acquire an anomalous dimension is the kinetic coefficient of the inertial term ($\sim\omega^2$), as we will further discuss below. This makes unviable  a fine compensation of the anomalous dimensions of the retarded  and Keldysh sectors, which would signal, whenever occurring, effective infrared thermalization~\cite{Sieberer2013,PhysRevB.94.085150}. Therefore, the RG fixed point discussed in the following explicitly breaks   fluctuation-dissipation relations and cannot have an equilibrium counterpart, distinctly from other instances of non-equilibrium open criticality~\cite{mitra06,PhysRevB.85.184302, DallaTorreDiehl2013, maghrebi2016nonequilibrium, foss2017emergent, kirton2019introduction}. We   now study   the  critical regime     $m\to0$ for $\alpha<1$. \\
 
 \paragraph{RG fixed point and criticality for $\alpha<1$ -- } 
 We will now complement   the Gaussian action in Eq.~\eqref{gauss} with non-linear terms    following the  canonical power counting   just discussed.
 For $\alpha<1$, we have $P^{R/A}\sim q^{2\alpha}$ and $P^K\sim q^\alpha$, with   canonical scaling dimensions for the classical and quantum fields $\varphi_c\sim q^{d/2-\alpha}$,~$\varphi_q\sim q^{d/2}$, and accordingly a lower critical dimension  of $d_l=2\alpha$. 
 Below the upper critical dimension $d_u=4\alpha$ the classical non-linear term $(u_c/4!)\varphi^3_c\varphi_q$ is relevant. The next RG leading non-linearity  appears at $d<3\alpha$ where the additive noise term $i(\kappa/2) \varphi^2_c\varphi^2_q$ and the sextic term $(\lambda/5!) \varphi^5_c\varphi_q$ are both RG relevant. Quantum vertices with higher powers of quantum fields are always irrelevant hinting at the semi-classical nature of the fixed point, and marking a difference with previous studies on quantum criticality induced by dark states~\cite{proseneisert,Moos,dalla2010quantum, Marino2016}. 
 Notice that similarly to the long-range interacting model A of the Halperin-Hohenberg classification~\cite{halimeh2021quantum} we have a dynamical critical exponent $z=\alpha$, but different lower and upper critical dimensions due to the gapless nature of the noise, suggesting that the scaling regime studied here belongs to a different universality class.  
 Similarly there are neat differences with the canonical power counting of the zero-temperature critical long-range   Ising model, where $z=\alpha/2$.

In order to find the interacting fixed point, we run a one-loop resummed RG on the effective potential  including     relevant non-linearities~\cite{amit2005field,peskin2018introduction,Sieberer2016review}. {Technical details are reported in~\cite{SM}}.  We employ a   sharp cutoff in momentum space $k<q<\Lambda$ where $k$ is the running RG scale  and $\Lambda$ an UV regulator. In the following we parametrize the flow of the couplings in terms of the  RG time $t=\ln k$. 
 We first consider a leading order $\epsilon$-expansion, right below the upper critical dimension $\epsilon\equiv d_u-d\ll1$ (where $d_u=4\alpha$). We follow the canonical rescaling discussed above, $\tilde{m}\sim m/k^{2\alpha}$, $\tilde{\Gamma}_0\sim \Gamma_0/(\Gamma_1k^\alpha)$ and $\tilde{u}_c\sim u_c/k^{4\alpha-d}$, and we find from the following rescaled beta functions 
\begin{equation}\label{beta}
\begin{split}
&\partial_t\tilde{m}=-2\alpha\tilde{m}+\frac{\tilde{u}_c(-2\tilde{m}+(1+\tilde{\Gamma}_0)^2)}{4(1+\tilde{\Gamma}_0)^6},\\ 
&\partial_t\tilde{u}_c=-\epsilon\tilde{u}_c-\frac{3\tilde{u}^2_c}{2(1+\tilde{\Gamma}_0)^6}, \quad \partial_t\tilde{\Gamma}_0=-\alpha\tilde{\Gamma}_0,
\end{split}\end{equation}
a  Wilson-Fisher (WF) fixed point at $(\tilde{m}^*,\tilde{\Gamma}^*_0, \tilde{u}^*_c)=(-\epsilon/(12\alpha), 0, -2\epsilon/3)$, with a correlation length critical exponent $\xi\sim m^{-\nu}$, $\nu=1/(2\alpha)+\epsilon/(12\alpha^2)$. This fixed point has an additional unstable RG direction corresponding to perturbations around the fixed point value $\tilde{\Gamma}^*_0=0$, in agreement with the requirement to fine  tune the Lindbladian gap ($\Gamma_0\to0$) in addition to the closing of the spectral one ($m\to0$).  This is   in full analogy with the RG relevance of   temperature at equilibrium quantum critical points, which is as well responsible for the onset of an additional  RG   unstable direction~\cite{Sondhi1997, vojta2003quantum, sachdev2007quantum}.
 At $\mathcal{O}(\epsilon^2)$ we find $z\simeq\alpha+\epsilon^2/(24(1+4\alpha^2))$ {following similar calculations performed for critical Langevin models~\cite{maghrebi2016causality,Tauber2014}}. 
 %

In Eqs.~\eqref{beta} the flow of $\tilde{\Gamma}_0$ is solely governed by its canonical dimension. To find a non-trivial WF fixed point for $\Gamma_0$,   we need   the multiplicative noise $i(\kappa/2) \varphi^2_c\varphi^2_q$ to be RG relevant. As discussed above, this occurs for $\alpha>d/3$, giving to the Gaussian noise sector, $P^K$,  a one-loop dressing proportional to $\sim \kappa\int_{Q}G^K(Q)$. %
For consistency with RG relevance  we have  also to include   the sextic vertex $\propto\lambda$ ({see ~\cite{SM} for details}). By evaluating the one-loop resummed RG flow  at $d=2$ and $\alpha=0.7$, we find a WF fixed point $(\tilde{m}^*,\tilde{\Gamma}^*_0, \tilde{u}^*_c, \tilde{\kappa}^*, \tilde{\lambda}^*)=(0.04, 0.23, 2.53, -1.98, -0.93)$ with still two unstable directions; the one associated to the spectral mass yields $\nu\simeq0.71$.
Loop corrections to $\Gamma_0$ in vicinity to this fixed point, renormalize the condition $\Gamma_0(k)\simeq \Gamma_1k^{\alpha}$ for suppression of the dark mode
from incoherent spontaneous emission. Following  a calculation  contained in   Ref.~\cite{PhysRevB.94.085150} ({summarized also in~\cite{SM}}), we find that at distances larger than  the inverse of $k^*\simeq 10^{-6}\Lambda_G$, the novel scaling is superseded by a conventional non-critical thermal Ising theory (as also mentioned above). Here $\Lambda_G$  is the so called  Ginzburg scale~\cite{amit2005field}: at distances larger than  $\Lambda^{-1}_G$, correlation functions scale universally with the critical exponents of the WF fixed point. For distances smaller than $\Lambda^{-1}_G$ correlation functions are instead dominated by non-universal corrections (lattice effects, RG irrelevant spin interactions, etc).    
%
%
{From the side of dynamics, upon initializing the  spin model \eqref{Ising}-\eqref{eq:model} sufficiently away from the eventual    steady state, it will enter, after a transient ($t\lesssim \Lambda^{-1}_G$), into a self-similar scaling regime where spatial- and time-resolved spin correlations are governed by the critical exponents of the WF fixed point. Such dynamical scaling regime persists  until spontaneous emission will 'heat' the  dark modes at times   larger than the inverse of $k^*$; at these times, the critical long-wavelength theory will crossover into a conventional Langevin theory.}  \\

 \paragraph{Fixed point for $\alpha>1$ -- } By inspection of Eqs.~\eqref{eqP} we notice that for $\alpha>1$ the kinetic coefficient $\sim Kq^2$ in the advanced/retarded sector dominates over the $\propto \Gamma^2_1 q^{2\alpha}$ term resulting from non-local dissipation. This leads to a dynamical critical exponent $z=1$, with the term $\propto \omega \Gamma_q$ now negligible in the infrared; in other words, we have an Ising model with short range interactions   and a $\propto \Gamma_1q^\alpha$ Markovian noise.
 This changes the critical properties of the theory as summarized in Table~\ref{tabella}. At this WF fixed point quantum terms such as the quartic $u_q\varphi^3_q\varphi_c$ are irrelevant, unless $\alpha>2$. However, as $\alpha$ increases the spatial support of losses quickly shrinks~\cite{kushal}, retrieving uninteresting   local dissipation effects similar to $\Gamma_0$.\\
 
  \paragraph{Competing long-range interactions and losses -- } Finally, we consider the scenario where long-range Ising interactions, $\sum_{\langle i,j\rangle}\frac{J}{|i-j|^{1+\alpha}} \sigma^x_i \sigma^x_j$, compete with   'long-range' losses. 
  Such term adds a $Jq^\alpha$ contribution to $P^{R/A}$~\cite{fisher1972critical,dutta2001phase,vodola2014kitaev, maghrebi2016causality}.
  %
  %
By inspection of Eqs.~\eqref{eqP}, we realize that   for $\alpha>2$ we recover the same scaling discussed above for   'long-range' losses with $\alpha>1$.  For $\alpha<2$, instead, we   find a leading scaling   $P^{R/A}\sim Jq^\alpha$, since long-range interactions suppress at small momenta the contribution of non-local losses in the spectral sector ($z=\alpha/2$). This is equivalent to the critical scaling of a long-range interacting Ising model driven by   a $\propto \Gamma_1q^\alpha$ Markovian noise, and it is a limit where classical and quantum vertices scale alike, $u_{c,q}\sim q^{3\alpha/2-d}$. Such quantum scaling regime is formally equivalent to a critical zero-temperature long-range interacting Ising model~\cite{fisher1972critical,dutta2001phase}. The associated critical dimensions and exponents are summarized in   Table~\ref{tabella} ({they do not hold for  $d=3$}).

\paragraph{Perspectives -- }   
%
%
{A recent  cavity QED experiment~\cite{periwal2021programmable}  demonstrates the tunability of non-local spin couplings, suggesting  that  the exploration of     programmable non-unitary interactions~\cite{kushal} in critical  spin systems may belong to near-term implementations. }
{An interesting follow-up research direction   would consist} in focusing on richer driven-open platforms, where incoherent losses/pumps and dephasing channels with non-local spatial character can compete. For instance, revising  driven-open condensates with   an $O(2)$ order parameter~\cite{Sieberer2013, Sieberer2014,altman2015two,Marino2016,zamora2017tuning}   appears  a natural  perspective.
{In the same spirit, the effective field theory derived in this work can be considered as a natural starting point for an extension to models with different symmetries or equipped with global conservation laws, in the pursuit of an Halperin-Hoenberg~\cite{Hohenberg1977} classification  of critical theories with tunable dark states.}
{It  also appears important to access quantitatively the value of the critical exponents (and the radius of convergence of the $\epsilon$-expansion) using methods, like functional-RG,  which are technically  suited to perform   loop resummations in models with long-range interactions~\cite{defenu2015fixed}. 
However, the   $\epsilon$-expansion of our work is expected to describe   critical properties at least qualitatively, as it also occurs   in hermitian long-range    Ising models~\cite{dutta2001phase, maghrebi2016causality, defenu2015fixed}, or as it would be for the  large-$N$ version~\cite{Moshe2003}  of the field theory~\eqref{gauss}.
In all these respects, our findings can be regarded as a seed for   technically richer  explorations. }

{Finally, we believe it would be extremely interesting to study the effect of long-range losses on the paramagnet and ferromagnet separated by the critical point. This appears, however, as a technically challenging task since it requires to solve the non-diagonal Liouvillian in~\eqref{eq:model} beyond semi-classical limits where its many-body dynamics have been  efficiently simulated so far~\cite{seetharam2021dynamical}. Quantum kinetic equations based on  Majorana fermions representation of spins~\cite{Demler2015} could represent a possible avenue to find correlations in this case.}
\\ 

\paragraph{Acknowledgements} We acknowledge N. Defenu, E. Fradkin, M. Maghrebi, A. Pi\~neiro-Orioli, K. Seetharam  for useful discussions and  S. Diehl, S. Kelly for proof reading the manuscript. 
This project has been supported by the Deutsche Forschungsgemeinschaft (DFG, German Research Foundation), Project-ID 429529648, TRR 306 QuCoLiMa ('Quantum Cooperativity of Light and Matter') and   in part by the National Science Foundation under Grant No. NSF PHY-1748958 (KITP program 'Non-Equilibrium Universality: From Classical to Quantum and Back').

\bibliography{biblio}
\bibliographystyle{apsrev4-1}

\end{document}